# Track-Consistency-Based GNSS RFI Monitoring Using Crowdsourced ADS-B Sensor Networks

Sanghyeon Park[1], Halim Lee[2], and Pyo-Woong Son[1*]

[1] Department of Electronics Engineering, Chungbuk National University, Cheongju 28644, South Korea

[2] Department of Aeronautics and Astronautics, Stanford University, Stanford, CA 94305, USA

**Abstract—**Growing reports of global navigation satellite system (GNSS) radio-frequency interference (RFI) highlight the need for scalable wide-area sensing for situational awareness. Crowdsourced Automatic Dependent Surveillance–Broadcast (ADS-B) receiver networks form a large-scale opportunistic sensor network for GNSS RFI monitoring, but ADS-B quality indicators may remain high during abnormal reported-position behavior, and heterogeneous receiver timestamping can produce apparent speed spikes. This letter proposes a three-stage framework that screens position-jump candidates, verifies local track consistency to suppress timing artifacts, and groups confirmed anomalies into traffic-adaptive multi-aircraft events. Using 607 million 1090-MHz ADS-B reports over Northeast Asia from December 2025 to February 2026, the framework identified 250 event clusters within the validity window of Notice to Airmen (NOTAM) RKRR Z1401/25 and none in the pre-NOTAM period. During the NOTAM-valid period, more than 99% of confirmed anomalies remained in high quality-indicator regimes, suggesting that track-consistency verification provides a complementary sensing criterion for GNSS RFI monitoring.

**Keywords—**ADS-B, crowdsourced sensor networks, GNSS RFI, spatiotemporal clustering, track-consistency verification.

## I. INTRODUCTION

Reports of global navigation satellite system (GNSS) radio-frequency interference (RFI), including jamming and spoofing [1], have increased worldwide, particularly in conflict-affected regions [2], motivating wide-area airspace monitoring. Crowdsourced Automatic Dependent Surveillance–Broadcast (ADS-B) reception networks provide an opportunistic sensing modality for wide-area GNSS RFI monitoring. Civil aircraft broadcast global positioning system (GPS)-derived positions, aircraft-state information, and quality indicators to distributed receivers; interference-induced positioning errors can therefore appear in ADS-B-reported positions. From a sensor-network perspective, aircraft act as mobile GNSS sensor nodes within a large-scale distributed sensing infrastructure.

Existing ADS-B-based GNSS RFI monitoring methods can be broadly divided into three approaches. The first approach infers potential GPS degradation from ADS-B quality indicators, such as the Navigation Integrity Category (NIC) and the Navigation Accuracy Category for Position (NACp) [3]–[7]. These indicators have been widely used as proxies for positioning performance degradation. However, our dataset includes cases in which reported positions deviate abruptly from the local aircraft track while NIC and NACp remain high, motivating a complementary check of the reported position against the local track of the same aircraft. The second approach detects position-jumps in ADS-B-reported positions and applies spatial clustering to identify regional anomaly patterns [8]. However, receiver clocks in crowdsourced ADS-B networks exhibit heterogeneous drift and require continual synchronization [9]; in position-jump screening, such timestamp inconsistencies can manifest as apparent speed spikes even when broadcast positions remain locally consistent with the track, motivating a local track-consistency check on each candidate before event clustering. The third approach applies trajectory-level consistency to individual tracks [10], [11], but does not address the receiver-timing artifacts specific to crowdsourced reception. In addition, applying a single density threshold across all regions can be sensitive to spatially uneven air traffic, motivating a traffic-adaptive density

Corresponding authors: Halim Lee and Pyo-Woong Son (e-mail: halimlee@stanford.edu; pwson@cbnu.ac.kr).

requirement for event clustering.

Table 1. Capability comparison with representative related ADS-B-based approaches.

| Related approaches | Timing artifacts | Track verification | Multi-aircraft events | Traffic-adaptive |
|---|---|---|---|---|
| Quality-indicator studies [3]–[7] | ✗ | ✗ | ✗ | ✗ |
| Position-jump and clustering study [8] | ✗ | ✗ | ✓ | ✗ |
| Trajectory-consistency studies [10], [11] | ✗ | ✓ | ✗ | ✗ |
| **This work** | ✓ | ✓ | ✓ | ✓ |

These blind spots motivate a framework that verifies position-jump candidates against neighboring reports of the same aircraft before traffic-adaptive multi-aircraft clustering, complementing NIC/NACp screening while suppressing timing-induced candidates. Table 1 summarizes how this combination differs from existing approaches. Using approximately 607 million OpenSky ADS-B reports [9] collected over three months in Northeast Asia, we show that the complete framework exhibits stronger temporal consistency with the validity period of GNSS-related Notice to Airmen (NOTAM) RKRR Z1401/25 [12] than the prior position-jump clustering method [8].

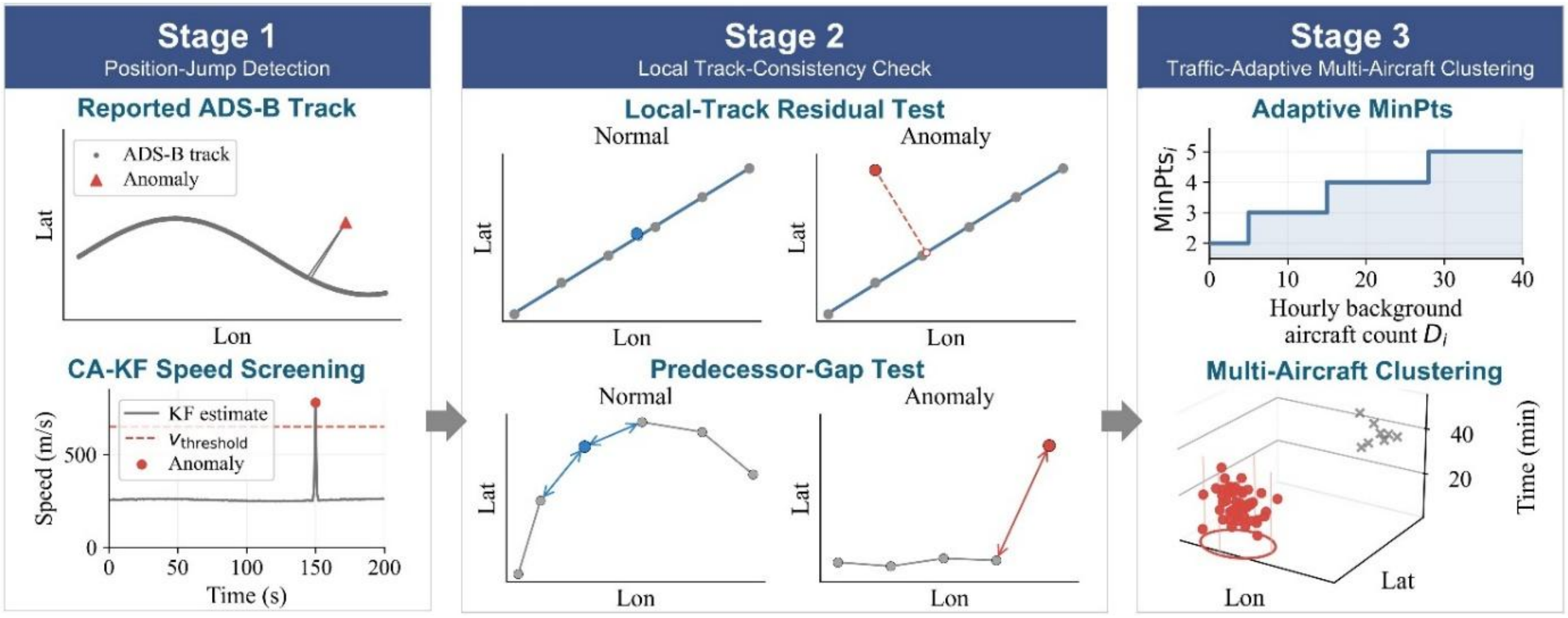

Fig. 1. Three-stage ADS-B-based GNSS RFI detection framework.

## II. PROPOSED FRAMEWORK

The proposed framework processes raw ADS-B reports into regional anomaly events through three stages, as shown in Fig. 1. Each ADS-B report from aircraft $j$ is denoted by

$$m_{j,k} = \left(t_{j,k},\ \varphi_{j,k},\ \lambda_{j,k},\ h_{j,k},\ \text{NIC}_{j,k},\ \text{NACp}_{j,k}\right), \tag{1}$$

where $k$ indexes successive reports, $t_{j,k}$ is the reception timestamp, and $(\varphi_{j,k}, \lambda_{j,k}, h_{j,k})$ are the reported latitude, longitude, and altitude. For distance calculations, the horizontal position $(\varphi_{j,k}, \lambda_{j,k})$ is projected onto a local Cartesian coordinate $p_{j,k}$. The framework first forms a position-jump candidate set $\mathcal{C}$, then retains a subset $\mathcal{A}$ of confirmed position anomalies, and finally groups $\mathcal{A}$ into multi-aircraft event clusters.

### A. Stage 1: Position-Jump Detection

For each aircraft, a constant-acceleration Kalman Filter (CA-KF) [8] is applied to the time-ordered position sequence $\{p_{j,k}\}$, yielding an apparent horizontal speed estimate $\hat{v}_{j,k}$; reports whose estimated speed exceeds $v_{\text{th}}$ form the position-jump candidate set:

$$\mathcal{C} = \{\, m_{j,k} \colon \hat{v}_{j,k} > v_{\text{th}} \,\}. \tag{2}$$

Reports in $\mathcal{C}$ may include receiver-timing artifacts as well as actual spatial inconsistencies and are passed to Stage 2 for verification.

### *B. Stage 2: Local Track-Consistency Check*

Stage 2 verifies whether each candidate in $\mathcal{C}$ is inconsistent with neighboring reports from the same aircraft. A speed-based candidate can arise from an actual displacement or from receiver-timing errors that create an apparent speed spike without spatial inconsistency. To separate these cases, Stage 2 applies two sequential tests: a local-track residual test and a predecessor-gap test. The former evaluates geometric consistency with the surrounding track, whereas the latter evaluates sequential continuity with the immediately preceding report.

For each candidate $m_{j,k} \in \mathcal{C}$, a local context $\mathcal{N}_{\mathrm{j,k}}$ of $N$ neighboring reports of the same aircraft is restricted to a continuous track segment so that reports from different flight phases are not compared. Let $\Delta t_{j,k} = t_{j,k} - t_{j,k-1}$ denote the inter-report time gap; the segment satisfies

$$\Delta t_{j,k} \;\le\; \Delta t_{\max}\big(h_{j,k}\big), \tag{3}$$

where $\Delta t_{\max}\big(h_{j,k}\big)$ is the altitude-dependent maximum allowed inter-report gap, defined as

$$\Delta t_{\max}\big(h_{j,k}\big) = \begin{cases} \Delta t_{\text{low}}, & h_{j,k} < h_0 \\ \Delta t_{\text{high}}, & h_{j,k} \ge h_0 \end{cases}, \tag{4}$$

following [13], with the altitude threshold $h_0$ and gap limits listed in Table 2.

For candidate $m_{j,k}$, a two-dimensional cubic smoothing spline $\mathbf{s}_{j,k}^{(-k)}(u)$ is fitted to the local Cartesian positions in $\mathcal{N}_{j,k} \setminus \{m_{j,k}\}$, where $(-k)$ denotes leave-one-out (LOO) exclusion of the candidate. The parameter $u$ is the chord-length coordinate, defined as a cumulative Euclidean distance along the ordered neighboring reports. The local-track residual is the shortest distance from the candidate position to the fitted local reference track:

$$d_{j,k} = \min \;\| \, \mathrm{p}_{j,k} - \mathrm{s}_{j,k}^{(-k)}(\mathrm{u}) \, \|. \tag{5}$$

A small residual indicates that the candidate remains spatially compatible with the surrounding track despite its apparent speed spike. To set an adaptive local threshold, the same LOO procedure is applied to non-candidate reports in $\mathcal{N}_{\mathrm{j,k}}$, yielding a reference residual sample. Its log-normal upper-tail quantile gives

$$\tau_{\text{track},j,k} = \exp\big(\hat{\mu}_{j,k} + z_{1-\alpha}\,\hat{\sigma}_{j,k}\big), \tag{6}$$

where $\hat{\mu}_{j,k}$ and $\hat{\sigma}_{j,k}$ are estimated from the logarithmic reference residuals. The log-normal form was selected by Akaike Information Criterion (AIC) comparison with Weibull and Gamma alternatives. The candidate is treated as a local-track outlier when

$$d_{j,k} > \tau_{\text{track},j,k}. \tag{7}$$

The predecessor-gap test measures the displacement between the candidate and its immediate preceding report:

$$g_{j,k} = \| \, \mathrm{p}_{j,k} - \mathrm{p}_{j,k-1} \, \|. \tag{8}$$

A small $g_{j,k}$ indicates that the candidate remains spatially connected to the preceding report, so the large local-track residual is not sufficient evidence of an abrupt spatial displacement. The gap threshold $\tau_{\text{gap},j,k}$ is obtained from the local reference distribution of predecessor gaps in $\mathcal{N}_{\mathrm{j,k}}$ using the rule in (6), and the candidate satisfies the condition when $g_{j,k} > \tau_{\text{gap},j,k}$.

A candidate is retained for event-level analysis only when it satisfies both the local-track residual and predecessor-gap conditions:

$$\mathcal{A} = \big\{\, m_{j,k} \in \mathcal{C} \colon d_{j,k} > \tau_{\text{track},j,k} \;\wedge\; g_{j,k} > \tau_{\text{gap},j,k} \,\big\}. \tag{9}$$

### *C. Stage 3: Traffic-Adaptive Multi-Aircraft Clustering*

Stage 3 groups the retained anomalies in $\mathcal{A}$ into regional multi-aircraft events that recur within the same airspace and time window. Spatio-Temporal Density-Based Spatial Clustering of Applications with Noise (ST-DBSCAN) is applied to $\mathcal{A}$ with horizontal, vertical, and temporal radii $\varepsilon_s$, $\varepsilon_v$, and $\varepsilon_t$, selected from $k$-distance curves using the Kneedle criterion [14]. For each confirmed anomaly $a_n \in \mathcal{A}$ with position $\mathrm{p}_n$, altitude $h_n$, and timestamp $t_n$, the spatiotemporal neighborhood is defined as

$$\mathcal{B}(a_n) = \{\, a_m \in \mathcal{A} \colon \| \, \mathrm{p}_n - \mathrm{p}_m \, \| \le \varepsilon_s \;\wedge\; |h_n - h_m| \le \varepsilon_v \;\wedge\; |t_n - t_m| \le \varepsilon_t \,\}, \tag{10}$$

where $(p_m, h_m, t_m)$ are defined analogously for $a_m \in \mathcal{A}$. The anomaly $a_n$ is treated as a core point when

$$|\mathcal{B}(a_n)| \geq \text{MinPts}. \quad (11)$$

Here, MinPts denotes the minimum number of confirmed anomalies required within $\mathcal{B}(a_n)$ for $a_n$ to be designated a core point. Because air traffic density varies spatially, we adapt $\text{MinPts}_i$ to the local background traffic via a Poisson tail, requiring stronger multi-aircraft support in dense regions and fewer supporting aircraft in sparse regions. For each confirmed anomaly $a_n \in \mathcal{A}$ located in grid cell $i$, let $D_i$ denote the hourly background aircraft count of the cell. The expected number of background-supported anomalies within ST-DBSCAN search window is modeled as

$$\mu_i = D_i \cdot S \cdot c, \; S = (\pi \, \varepsilon_s^2 / A_{\text{grid}}) \cdot (\varepsilon_t / T_{\text{grid}}), \quad (12)$$

$$\text{MinPts}_i = \max\{\text{MinPts}_{\text{base}}, \; F_{\text{Poisson}}^{-1}(1-\alpha; \mu_i) + 1\}, \quad (13)$$

where $F_{\text{Poisson}}^{-1}(\cdot\,; \mu_i)$ denotes the inverse cumulative distribution function of a Poisson distribution with mean $\mu_i$, $\alpha$ is the Poisson tail, $\text{MinPts}_{\text{base}}$ is the minimum allowed neighborhood count, and $c$ is a calibration factor that converts the local background traffic level into an expected anomaly rate. Each final event cluster is additionally required to contain at least two unique aircraft.

## III. EXPERIMENTAL VALIDATION

### A. Dataset and Parameter Settings

ADS-B reports on the 1090-MHz extended-squitter (1090ES) channel were collected from the OpenSky Network over Northeast Asia and adjacent airspace (27°–46°N, 118°–139°E) from December 2025 through February 2026, totaling 606,922,091 reports from 8,015 unique aircraft. Table 2 summarizes the data-derived and literature-based parameters; the remaining CA-KF settings follow [8].

Table 2. Parameter settings for the proposed framework.

| Symbol | Value | Symbol | Value |
|---|---|---|---|
| $v_{\text{th}}$ | 650 m/s [8] | $\varepsilon_{\text{s}}$ | 35 km |
| $N$ | 100 | $\varepsilon_{\text{t}}$ | 75 min |
| $h_0$ | 1,000 m [13] | $\varepsilon_{\text{v}}$ | 2.5 km |
| $\Delta t_{\text{low}}$ | 10 min [13] | Grid Resolution | 0.1 ° |
| $\Delta t_{\text{high}}$ | 30 min [13] | $A_{\text{grid}}$ | 100 $\text{km}^2$ |
| $c$ | $1.08\times10^{-3}$ | $T_{\text{grid}}$ | 60 min |
| $\alpha$ | 0.01 | $\text{MinPts}_{\text{base}}$ | 2 |

### B. NOTAM-Window Analysis

Because the exact timing and source of GNSS RFI activity are generally not directly observable, we use NOTAM RKRR Z1401/25 as an external temporal reference rather than an event-level ground-truth label. The NOTAM advised pilots that GPS signals had been reported as unreliable or intermittently lost in the Incheon Flight Information Region (FIR) [12]. Based on its validity period, the dataset is partitioned into pre-NOTAM (1–28 December 2025), NOTAM-valid (29 December 2025–31 January 2026), and post-NOTAM (1–28 February 2026) periods.

To compare temporal consistency with the NOTAM-valid window, we applied both the prior ADS-B position-jump clustering method [8] and the proposed framework to the same Stage 1 position-jump candidates in each period (Table 3). The proposed framework produced 250 event clusters involving 817 unique aircraft during the NOTAM-valid period and no clusters in the pre-NOTAM period. In contrast, the prior method [8] produced 56, 47, and 30 clusters across the three periods, indicating weaker temporal consistency with the NOTAM validity window.

Using NIC ≥ 7 and NACp ≥ 8, corresponding to the minimum ADS-B Out containment and accuracy requirements of §91.227 [15] and to the indicator behaviour reported in [6], 99.54% and 99.68% of the 3,065 confirmed position anomalies during the NOTAM-valid period remained in these high-indicator regimes (Table 3) and would therefore have been missed by NIC/NACp threshold-based screening. Table 4 shows the stage-wise reduction of the candidate set: the largest reduction occurs at Stage 2a (local-track residual), which excludes more than 93% of the position-jump candidates in all periods, while Stage 2b further isolates reports with both inconsistencies for clustering.

Table 3. Temporal consistency with NOTAM RKRR Z1401/25. Results are reported as event clusters/unique aircraft; NIC-/NACp-high ratios are computed over confirmed anomalies of the proposed framework.

| Period | Pre | Valid | Post |
|---|---|---|---|
| Date | 12/01–12/28 | 12/29–01/31 | 02/01–02/28 |
| NOTAM Z1401/25 | Not valid | Valid | Not valid |
| NIC-high ratio | 76.92% | 99.54% | 96.80% |
| NACp-high ratio | 100.00% | 99.68% | 100.00% |
| Prior method [8] | 56 / 1,573 | 47 / 2,871 | 30 / 1,417 |
| Proposed | 0 / 0 | 250 / 817 | 19 / 68 |

Table 4. Stage-wise processing cascade across the three analysis periods. Entries are count/aircraft.

| Period | Pre | Valid | Post |
|---|---|---|---|
| 0. Raw messages | 232,526,676 / 6,210 | 211,726,476 / 6,402 | 162,668,939 / 5,969 |
| 1. Position-jump detection | 4,836 / 1,652 | 54,309 / 2,932 | 5,500 / 1,536 |
| 2a. Local-track residual test | 4,767 / 1,640 | 50,510 / 2,904 | 5,263 / 1,507 |
| 2b. Predecessor-gap test | 4 / 2 | 734 / 525 | 18 / 18 |
| 2c. Confirmed anomalies | 65 / 60 | 3,065 / 1,228 | 219 / 187 |
| 3. Event clusters | 0 / 0 | 250 / 817 | 19 / 68 |

### *C. Case Study: 5 January 2026*

On 5 January 2026, the framework identified a multi-aircraft event cluster centered near (37.28°N, 127.06°E) with a spatial radius of 32.45 km, where 6 of 14 transiting aircraft (42.9%) showed confirmed position anomalies spanning altitudes of 6.5–10.4 km (Fig. 2). The impact was not uniform across aircraft, which may reflect aircraft-dependent differences in onboard GNSS configurations [6].

The affected aircraft exhibit elevated local-track residual and predecessor-gap values near the event region (Fig. 3), whereas most background aircraft remain below the corresponding local thresholds. Each aircraft shows a track-consistency violation after entering the same spatial region, supporting an airspace-localized interpretation rather than an isolated single-aircraft failure.

In a representative single-aircraft case, both NIC and NACp remained high throughout the event (Fig. 4), yet the proposed track-consistency check still confirmed position anomalies in this aircraft.

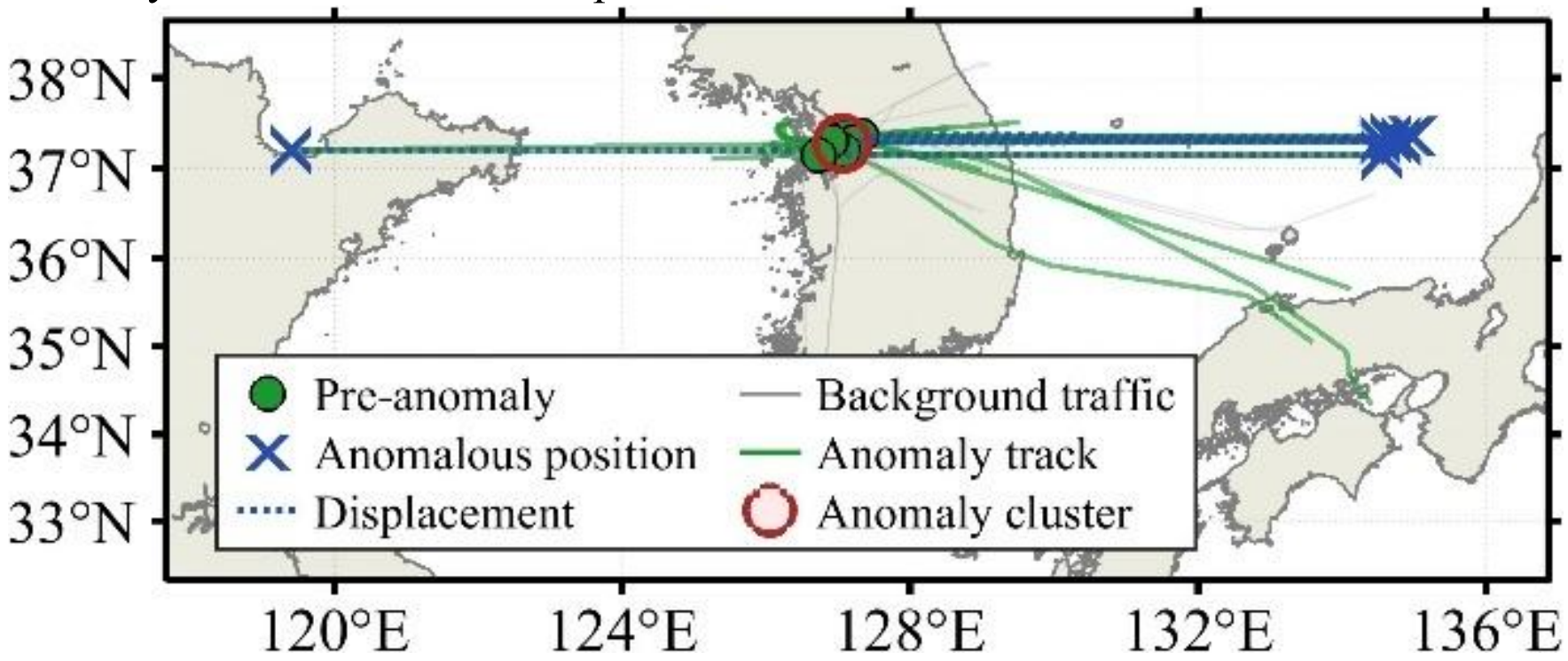

Fig. 2. Event cluster detected on 5 January 2026 for six affected aircraft near (37.28°N, 127.06°E).

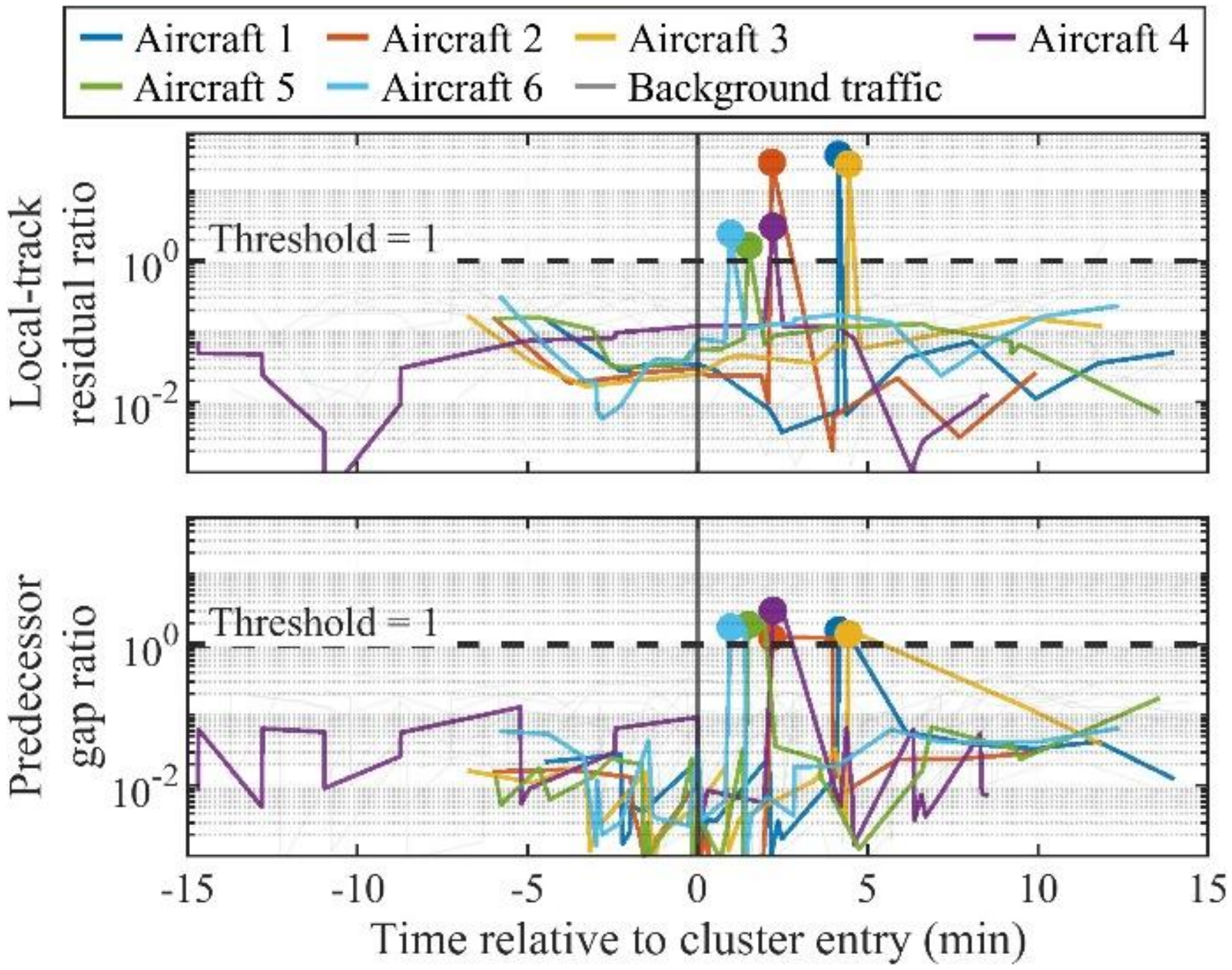

Fig. 3. Threshold-normalized local-track residual and predecessor gap of the affected aircraft, aligned to cluster entry (t = 0); background traffic in gray, and the dashed line denotes the unit threshold. A report is confirmed only when both panels exceed unity.

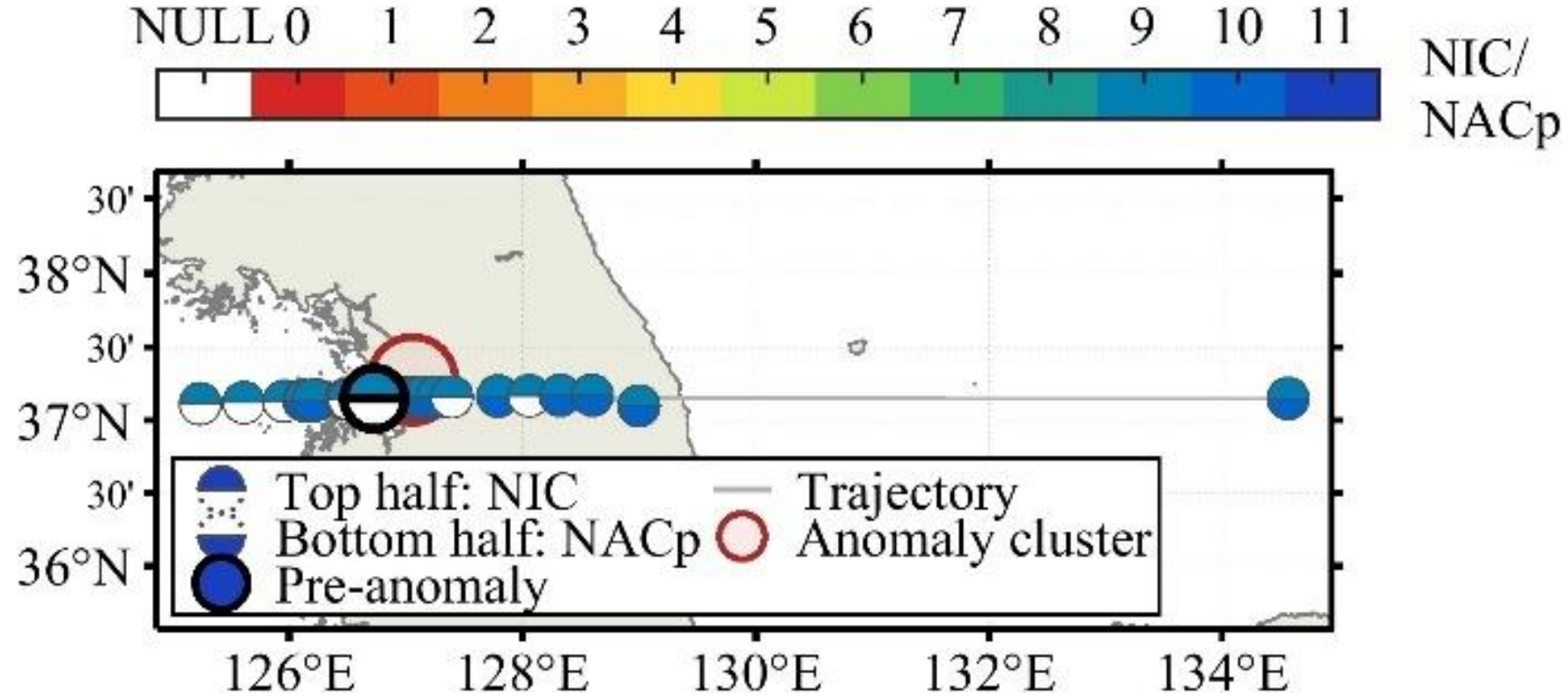

Fig. 4. Representative single-aircraft view showing confirmed position anomalies despite high NIC and NACp values.

## IV. CONCLUSION

This letter has proposed a three-stage framework for identifying GNSS-RFI-consistent anomaly events that manifest as abrupt inconsistencies in crowdsourced 1090ES ADS-B reported positions. In the analyzed three-month dataset, 250 event clusters were concentrated within the validity period of NOTAM RKRR Z1401/25 over the Incheon FIR, whereas no event cluster appeared in the pre-NOTAM reference period and the prior method produced clusters in all three periods. Because 99.54%/99.68% of the confirmed anomalies during the NOTAM-valid period remained in NIC-/NACp-high regimes, track-consistency verification can provide a complementary sensing criterion beyond ADS-B quality indicators. Independent GNSS-monitoring measurements could further corroborate event timing.

## ACKNOWLEDGMENT

This work was supported in part by the Korea Agency for Infrastructure Technology Advancement (KAIA) grant funded by the Ministry of Land, Infrastructure and Transport (Grant No. RS-2026-25527966), and in part by the Regional Innovation System & Education (RISE) program through the Chungbuk Regional Innovation System & Education Center, funded by the Ministry of Education (MOE) and Chungcheongbuk-do, Republic of Korea (Grant No. 2025-RISE-11-014-03), and in part by Chungbuk National University Glocal30 project (2025).